\documentclass[11pt]{article}
\usepackage{a4,epsfig}

\usepackage[round,colon]{natbib}
\renewcommand{\cite}{\citep}
\newcommand{\be}{\begin{equation}}
\newcommand{\ee}{\end{equation}}
\newcommand{\bea}{\begin{eqnarray}}
\newcommand{\eea}{\end{eqnarray}}

\newcommand{\uu}{{\mathbf u}}
\newcommand{\EE}{\mbox{\sffamily\bfseries E}}
\newcommand{\CC}{\mbox{\sffamily\bfseries C}}
\newcommand{\KK}{\mbox{\sffamily\bfseries K}}
\newcommand{\QQ}{\mbox{\sffamily\bfseries Q}}
\newcommand{\sig}{\mbox{\boldmath $\sigma$}}
\newcommand{\bd}{\mbox{\boldmath$\cdot$}}
\newcommand{\T}{^{\rm T}}
\newcommand{\mident}{\mbox{\boldmath$1$}}
\newcommand{\gamdot}{\dot\gamma}

\def\(#1){(\ref{#1})}

\begin{document}

\title{Tensorial Constitutive Models for Disordered Foams, Dense Emulsions, and other Soft Nonergodic Materials}

\author{ 
M. E. Cates\\School of Physics, University of
Edinburgh,\\JCMB, King's
Buildings,\\ Mayfield Road, Edinburgh, EH9 3JZ, UK.
\and 
P. Sollich\\Department of Mathematics, King's College,\\ University of
London, Strand, London,
WC2R 2LS, UK.
}

\date{2 May 2003}

\maketitle
\begin{abstract}

In recent years, the paradigm of `soft glassy matter' has been used to
describe diverse nonergodic materials exhibiting strong local disorder
and slow mesoscopic rearrangement. As so far formulated, however, the
resulting `soft glassy rheology' (SGR) model treats the shear stress
in isolation, effectively `scalarizing' the stress and strain rate
tensors. Here we offer generalizations of the SGR model that combine
its nontrivial aging and yield properties with a tensorial structure
that can be specifically adapted, for example, to the description of
fluid film assemblies or disordered foams.

\end{abstract}

\section{Introduction}
Many soft materials, including foams, dense emulsions, slurries,
pastes, and textured morphologies of liquid crystals, are
characterized by the presence of structural disorder on a mesoscopic
scale (nanometres to microns), causing metastability and slow
dynamical evolution. Such materials are nonergodic, and can therefore
be viewed, in at least one sense, as glasses. In the current work we
address only shear-thinning materials (for generalizations to shear
thickening see \citet{HeaAjdCat01,HeaAjdCat02}) where data for the
steady state flow curve 
are very often fitted either to (a) the `Herschel-Bulkley' form,
$\sigma(\dot\gamma)-\sigma_Y \sim \dot\gamma^p$, where $\sigma_Y>0$ is
a yield stress and $0<p<1$;  or (b) to the `power law fluid' form
which is the same except with $\sigma_Y = 0$. 
(A survey of edible soft matter \cite{Holdsworth93} lists a very large
number of instances of such fits in the literature.) The linear
rheological spectra are also often close to power law fluid form, or
else exhibit near constant storage modulus $G'(\omega)$ and
anomalously flat or even rising loss modulus $G''(\omega)$ as
frequency is decreased. In many instances, aging effects are seen
\cite{HohCohAsn99,CloBorLei00,CohHoh01,ViaLeq02,ViaJurLeq02,CloBorMonLei03} in which the
material gradually gets 
more elastic and less lossy as time goes by. These effects can be quite
complicated, with the rate of aging depending on stress
\cite{CloBorLei00}.

\section{The SGR Model}
Much of the aforementioned phenomenology, including some but not all
aspects of the observed aging behavior, are captured by a simple and
generic model called the SGR (or `soft glassy rheology') model
\cite{SolLeqHebCat97,Sollich98,FieSolCat00}. This model is based on Bouchaud's
trap model of glasses \cite{Bouchaud92,BouComMon95}, and envisages mesoscopic
elements whose dynamics consists of independent hopping between local
traps (or free energy minima). In the context of e.g.\ a foam, such
`hopping' events correspond to yielding, where a cluster of
bubbles rearranges into a new and more energetically favourable
topological structure.
The hopping is controlled by an
effective temperature parameter $x = T_{\rm eff}/T_{\rm g}$
which lies near a glass transition ($x \simeq 1$). Such a transition
exists if, as we shall assume, the distribution of trap depths is
exponential. This choice allows many-body effects (which are
undoubtedly present near any glass transition) to be ignored without
losing the transition altogether, and is therefore an intrinsic part
of the trap model picture as usually formulated.

In its original form \cite{SolLeqHebCat97,Sollich98} which addresses only
simple shear flows, the SGR model combines the hopping dynamics of
Bouchaud's model with the buildup of local elastic shear strains in
the mesoscopic elements; these are assumed, between hops, to be
linearly elastic and to deform affinely with the applied flow. Upon
hopping, the local strain $l$ is reset to zero. The stored elastic
energy density $kl^2/2$ (with $k$ an elastic modulus) in each element is
offset against the trap depth and lowers the local activation barrier
to hopping; this leads to shear thinning. The dynamics of the SGR
model is contained in the time evolution equation for the probability
distribution $P(E,l,t)$ for a mesoscopic element being in a trap of
depth $E$ ($>0$) with local shear strain $l$:
\begin{equation}
\dot P(E,l,t) = -\dot\gamma\,\partial P/\partial l -\Gamma_0
e^{-(E-kl^2/2)/x} P(E,l,t) + \Gamma(t)\rho(E)\delta(l)
\label{sgreom}
\end{equation}
Here $\Gamma_0$ is an intrinsic jump rate, and $\Gamma(t) = \int
\Gamma_0 e^{-(E-kl^2/2)/x} P(E,l,t)\, dE\, dl$ is the overall jump
rate allowing for the modulation of barriers by the strain; $\rho(E) =
\exp(-E)$ denotes the distribution of trap depths into which elements
can jump. Note that in proper dimensional units, we should write
$(E-vkl^2/2)/(xk_{\rm B}T_{\rm g})$ instead of $(E-kl^2/2)/x$, with $v$
the volume of an element. In the following, we choose to measure
energy densities such as $kl^2/2$ in units of $k_{\rm B}T_{\rm g}/v$, and
energies such as $E$ in units of $k_{\rm B}T_{\rm g}$, so that the extra
dimensional factors disappear.
The macroscopic shear stress is taken to obey
\begin{equation}
\sigma(t) = \int P(E,l,t)\,kl \,dE\, dl.
\label{sgrstress}
\end{equation}

For a detailed discussion of how the aforementioned flow and aging
phenomenology arises from this SGR model, see
\citet{SolLeqHebCat97,Sollich98,FieSolCat00}. A review which puts this in a
broader context is \citet{Cates_LesHouches03}. There are, of course, many open
issues with the model. One of these concerns the interpretation of the
noise temperature $x$ and whether or not this should depend on flow
history: in this paper we assume it does not. Also, the rheological
aging predictions of the model, though
surprisingly rich \cite{FieSolCat00,ViaLeq02,ViaJurLeq02}, do not
include all those found 
experimentally; see e.g. \citet{HohCohAsn99,CloBorLei00,CohHoh01}. Nonetheless, the model represents a
useful step towards understanding the rheology of materials that are
not time-translation invariant \cite{FieSolCat00}.

A separate shortcoming of the SGR model lies in its tensorial
simplicity. Because simple shear flow and linear local elasticity are
both assumed, no normal stresses can ever arise. The strain variable
$l$ in Eq.\ref{sgreom} is effectively a scalar, as is the macroscopic
shear stress $\sigma$. The form chosen for the shear thinning is also
quite restricted. In what follows we address these shortcomings by
showing how the SGR model can be `tensorialized' with minimal damage
to the appealing phenomenology that it contains. This allows
for various forms of nonlinear elasticity at the mesoscale, and
also lets us consider arbitrary deformation histories rather than just
simple shear.

Our starting point is the constitutive equation for the standard,
scalar SGR model, whose derivation \cite{Sollich98} we briefly recall
here. We imagine that
at the zero of time the sample is prepared in a known state that has,
for simplicity, all mesoscopic elements unstrained; this assumption
could be relaxed, but saves algebra. This state is characterized by
$P(E,l,0) = P_0(E)\delta(l)$.
The constitutive equation then reads
\begin{equation}
\sigma(t) = G_0(z_{t0})k\gamma(t) +
\int_0^{t} \Gamma(t') G_1(z_{tt'})\,k\!\left[\gamma(t)-\gamma(t')\right]dt'
\label{sgrce}
\end{equation}
Here $ G_0(z_{t0})$ is the fraction of elements present originally (at
$t=0$) surviving to time $t>0$. The factor $k\gamma(t)$ is the stress
they contribute. The integral is over elements
created at $t'<t$, with creation rate $\Gamma(t')$ and survival
probability $G_1(z_{tt'})$ to time $t$; each has been strained through
$\gamma(t) -\gamma(t')$ and contributes stress accordingly. The
following equations complete the prescription \cite{Sollich98}:
\begin{equation}
1 = G_0(z_{t0})+\int_0^{t'}\Gamma(t') G_1(z_{tt'})\,dt'
\label{norm}
\end{equation}
\begin{equation}
z_{tt'} = \int_{t'}^t
\exp\left(k\left[\gamma(t'')-\gamma(t')\right]^2/2x
\right) dt''
\label{efftime}
\end{equation}
\begin{equation}
G_0(z) = \langle e^{-z\Gamma_0\exp(-E/x)}\rangle_{P_0}\;\;\;\;\;;\;\;\;\;\;
G_1(z) = \langle e^{-z\Gamma_0\exp(-E/x)}\rangle_{\rho} \;.
\label{gdefs}
\end{equation}
Equation \ref{norm} fixes, by normalization, the jump rate
$\Gamma(t')$ required in Eq.\ref{sgrce}. Equation \ref{efftime}
defines an `effective time interval' $z$ between times $t$ and $t'$;
this is defined in such a way as to absorb the factor by which the
jump rate out of a trap is enhanced due to the presence of a
strain. In other words, within the scalar SGR model, the nonlinear
effect of shear can be viewed as `making the clock tick faster' for
any element that has accumulated  nonzero value of the local strain
$l$ since its creation. Indeed, this is the {\em only} nonlinearity in
the model, which is why Eq.\ref{gdefs} defines the survival functions
for each class of element exactly as one would in Bouchaud's model,
except that $z$ replaces $t$.

\section{Tensorial SGR Models}
We now turn to the main agenda, which is to relax the tensorially
naive assumptions of the SGR model as formulated thus far. Fortunately
the structure and interpretation of the scalarized SGR constitutive
equation, Eq.\ref{sgrce}, allow one very easily to construct various
tensorial descriptions based on the same underlying physics, but with
more specific and fully tensorial models for the local elastic and
yield behaviour of the mesoscopic elements. Because the connection
with Bouchaud's trap model remains intact, these models will retain
the interesting aging behavior and spectral phenomenology outlined in
the introduction. The same applies to flow curves although these will
be somewhat modified by any additional nonlinearities now introduced
at the mesoscale.

We present this first for a schematic dumbell based SGR model. A somewhat simpler class of models, closer in
spirit to the scalar SGR picture, is then suggested. Members of this
class adapted to describe the case of foams and dense emulsions, are
then studied in more detail.

\subsection{A Dumbell SGR Model}
We consider an ensemble of dumbell-like objects. Each dumbell is
characterized by an end-to-end vector $\uu \equiv u_\mu$ of length $u \equiv |\uu|$; the stress contributed by
such dumbells is deemed to be $\sigma_{\mu\nu} = nk\langle
u_\mu u_\nu\rangle_P$ where the average is over a distribution function
$P(E,\uu,t)$; $n$ is the density of dumbells and $k$ an elastic
constant. (The units of $k$ are, in this section, different from those
used previously; e.g.\ $ku^2/2$ has the dimensions of an
energy and is therefore assumed to be expressed in dimensionless multiples
of $k_{\rm B}T_{\rm g}$.)
Thus far, apart from the appearance of the $E$ variable,
this formulation resembles a first step towards the upper convected
Maxwell model \cite{Cates_LesHouches03}, where the dumbell dynamics consists of
two particles advected by the flow connected by a spring. (This is
often used as a model for sub-entangled polymers.) In the present
context we are not restricted to thinking of our dumbells as polymers;
they represent unspecified elastic objects, each of which is held in
place by its neighbors but can make discrete stochastic rearrangements
that relax its stress locally. 

SGR-like dynamics (quite unlike that of the Maxwell model) is now
introduced by assuming that each dumbell follows the flow affinely,
except that from time to time it makes a jump to a completely new
configuration, with a jump rate $\Gamma_0\exp[-(E-ku^2/2)/x]$. Here
$E$ is an energy barrier which is lowered by the stored elastic energy
$ku^2/2$. (One can imagine the dumbells as hooked into a network of
neighbors, which deforms affinely; but when a particular dumbell
becomes too elongated, its connections to the network will be more
likely to break.) After a jump, the dumbell is assigned a new yield
energy $E$ drawn from the usual prior distribution $\rho(E) = e^{-E}$. It is also
assigned at random a new value of $\uu$, drawn from the equilibrium
distribution at noise temperature $x$, $p_{\rm eq}(\uu) \sim
\exp(-k\uu\bd\uu/2x)$.

The resulting equation of motion, closely analogous to Eq.\ref{sgreom}, is
\begin{equation}
\dot P(E,\uu,t) = - (\partial P/\partial \uu)\bd\KK\bd\uu -
\Gamma_0 \exp[{-(E-ku^2/2)/x}] P + \Gamma(t)\rho(E)p_{\rm eq}(\uu)
\label{tsgreom}
\end{equation} 
with $\KK$ the rate-of-strain tensor and $\Gamma(t) =
\Gamma_0\langle\exp[-(E-ku^2/2)/x] \rangle_P$ the total jump rate.
A new feature is the appearance of $p_{\rm eq}(\uu)$ in the last term to
replace $\delta(l)$ in the scalar SGR model: this is the closest we
can get to the assumption, made there, of `zero strain' in new
mesoscopic elements without actually setting $\uu$ to zero (in which
case every dumbell would collapse to a point and never be stretched
under the affine flow). This choice will recover Boltzmann equilibrium
for the $\uu$ (at temperature $x$) in the absence of flow, which is
not true of the scalar model: but it is a moot point whether this {\em
should} be recovered, since $x$ is not a true temperature
\cite{SolLeqHebCat97,Sollich98}. An alternative choice, probably not much
different in practice, would be to use $4\pi p_{\rm eq}(\uu) =
\delta\left((\uu\bd\uu)^{1/2}-(3x/k)^{1/2}\right)$, corresponding to
the selection of new dumbells with random directions on a sphere of
radius $\langle u^2\rangle_{\rm eq}^{1/2} = 3x/k$. 

Now assume that at time zero a system is prepared by some definite
process (e.g.\ a quench) that gives a known initial distribution of
dumbell barrier heights and vectors $P(E,\uu,0) = P_0(E)p_0(\uu)$
(chosen factorable for simplicity).
Then, following the same arguments as lead to Eq.\ref{sgrce}, the
constitutive equation can be written (with an additional coefficient
$\alpha$ discussed below)
\bea
\alpha^{-1}\sigma_{\mu\nu}(t) &=& \langle G_0(z_{t0}(\uu))nk
(\EE_{t0}\bd\uu)_\mu
(\EE_{t0}\bd\uu)_\nu
\rangle_{p_0}
\nonumber\\
& &{}+{} \int_0^{t} \Gamma(t') 
\langle G_1(z_{tt'}(\uu))nk 
(\EE_{tt'}\bd\uu)_\mu
(\EE_{tt'}\bd\uu)_\nu
\rangle_{p_{\rm
eq}} dt'\,.\label{tsgrce}
\eea
Here $\EE_{tt'}$ is the deformation tensor between times $t'$ and $t$,
while $G_0, G_1$ are precisely as defined in Eq.\ref{gdefs};
the total jump rate $\Gamma(t)$ is found from
\begin{equation}
1 = \langle G_0(z_{t0}(\uu)\rangle_{p_0}+\int_0^{t'}\Gamma(t')
\langle G_1(z_{tt'}(\uu))\rangle_{p_{\rm eq}}dt'
\label{tnorm}
\end{equation}
and the `effective time' variable $z$, which is now explicitly
dependent on the end-to-end vector $\uu$ with which an element was
created, obeys
\begin{equation}
z_{tt'}(\uu) = \int_{t'}^t \exp\left(k
|\EE_{t''t'}\bd\uu|^2
/2x \right) \, dt''\;.
\label{tefftime}
\end{equation}

The above constitutive equation involves, as well as a tensorial
dependence on strain of the elastic stress and of the yield energy, an
extra layer of averaging over the $\uu_i$ variables describing the
end-to-end vectors with which dumbells were created after their most
recent jump. This makes its quantitative analysis (which must mainly
be done numerically from this point onwards) rather cumbersome and we
do not pursue it in this paper. Nonetheless, it is clear by the
construction of the model that its behavior in simple shear flows will
differ relatively slightly from the scalar SGR model; we should expect
Herschel-Bulkley ($x<1$) and power-law-fluid ($x>1$) flow curves with
power law $G^*(\omega)$ spectra at $x>1$ and aging behavior, very
similar to that described by \citet{FieSolCat00}, for $x\le 1$. But,
unlike the scalar version, the model is capable of nontrivial
predictions for normal stresses under shear and also for the
rheological and aging behavior in elongational and mixed flows.

The model just presented has some features in common with that of
\citet{MicAppMolKiePor01}. If the dumbells are thought of as polymer strands,
it could be used to represent physical gels with cross links having a
broad (in the model, exponential) distribution of activation
energies. As formulated so far, these energy barriers are lowered in
height by the full stored energy in a network strand between
links. This is surely an exaggeration, since in practice only a part
of that energy can be converted into the work of breaking a
link. However, to improve this correspondence one could choose $\alpha
>1$ in Eq.\ref{tsgrce}. In this case the
stored energy appearing in the
lowering of energy barriers is only a fraction $1/\alpha$ of the
elastic energy of the strand, whereas the full value is used to
calculate the macroscopic state of stress. (Equivalently,  the plateau
modulus in the model is now larger by a factor of $\alpha$.) 

\subsection{A Better Class of Model}
The extra averaging required above, in passing from the scalar to the
dumbell SGR model, is rather a nuisance. Arguably, though, it ought to
be redundant.  After all, the basic elements of the SGR picture were
not intended to represent individial polymers or particles (as the
dumbell idea tacitly assumes) but mesoscopic elements, within which a
degree of local averaging can already be presumed. One could thus hope
for a simpler description in which such elements have a distribution
of yield energies $E$, but otherwise are taken as isotropic bodies
with a well-behaved, deterministic elastic response, albeit in most
cases nonlinear. No averaging over $\uu_i$ would be necessary if we
replaced our dumbells with elastic spheres, for example. 

The general structure of the SGR-type constitutive equation that
results from this assumption of local averaging within a meso-element
is as follows:
\begin{equation}
\sig(t) = G_0(z_{t0})\QQ(\EE_{t0}) + \int_0^{t} \Gamma(t')
G_1(z_{tt'}) \QQ(\EE_{tt'})\,dt'
\label{tgrce2}
\end{equation}
\begin{equation}
z_{tt'} = \int_{t'}^t \exp\left[R(\EE_{t''t'})/x\right]dt''
\label{tefftime2}
\end{equation}
with $G_0, G_1$ and $\Gamma(t)$ obeying
Eqs.\ref{norm},\ref{gdefs}. Here $\QQ$ and $R$ are tensor and
scalar functions of $\EE$ that can be freely chosen. Suitable
choices for foams and dense emulsions are suggested below, but models
for other specific materials could employ quite different forms for
these. By construction, the pre-averaged tensorial models defined by
Eqs.\ref{tgrce2},\ref{tefftime2} are, for bland choices of $\QQ$ and
$R$, again expected to behave rather like the scalar SGR model (with
glass transition at $x=1$, aging and power-law fluid regimes etc.),
while at the same time describing nontrivial normal stress effects
under shear flow, and offering tractable models for the behavior of
soft glassy materials in extensional flow. 
Shear thickening models, along the lines developed from the scalar SGR model by
\citet{HeaAjdCat01,HeaAjdCat02}, could also be introduced with suitable
choices of 
$\QQ$ and $R$; in particular, if $\QQ$ is not strongly
strain-thinning, a choice of $R$ that drops sharply at intermediate
strains and rises again for large ones should lead to shear-thickening
(and perhaps static jamming) behavior \cite{HeaAjdCat01,HeaAjdCat02}. 

\subsection{Tensorial SGR Models for Foams and Dense Emulsions}
Turning now to foams and dense emulsions, what are suitable choices
for $\QQ$ and $R$? The first of these determines (to within a
factor of the plateau modulus $G$ which we set equal to unity
below) the stress created in a mesoscopic element as a result of a
deformation. In particular, the instantaneous response to a nonlinear
step strain is wholly controlled by $\QQ$. Therefore it fixes, for
example, the ratio $\varphi = N_2/N_1$ of first normal stress
differences in such an instantaneous response. Another significant
quantity, closely related to $\QQ$, is the ratio $\chi = G/{\cal
F}_0$ between the elastic modulus and the stored free energy density
${\cal F}_0$, which resides in the interfacial energy of the fluid
films in an unstrained state. On dimensional grounds, $G$ and ${\cal
F}$ are proportional; each scales as $\sigma/\xi$ where $\sigma$ is an
interfacial tension and $\xi$ a characteristic length scale associated
with a foam droplet.

Turning to the scalar quantity $R$, this represents the lowering of
rearrangement barriers due to stored strain energy. In principle the
lowering of a barrier could have a somewhat complicated dependence on
the deformation $\EE$ of the given element, but we assume for
simplicity that it depends only on the free energy density ${\cal
F}(\EE)$ in a mesoscopic element under strain. Moreover, again for the
sake of simplicity, we assume this dependence is linear:
\begin{equation}
R(\EE) = \lambda \frac{{\cal F}(\EE)-{\cal F}_0}{{\cal F}_0}
\label{barriergen}
\end{equation}
where $\lambda$ is a parameter that should be, with our choice of
units, of order one. (This parameter obviates the need for the
parameter $\alpha$ introduced earlier in Eq.\ref{tsgrce}; it plays the
same role of determining what fraction of the stored elastic energy
can be used as work to overcome a rearrangement barrier.) With $R$
defined as in Eq.\ref{barriergen}, the ratio $\chi$ can be read off
from an expansion for small shear strain $\gamma$ as
$R=\lambda\chi\gamma^2/2 + \cdots$.

We now introduce three possible models for the elasticity of fluid
film assemblies (foams and dense emulsions) that predict approximate
forms for both the step-strain response tensor $\QQ(\EE)$ and
the stored free energy ${\mathcal{F}}(\EE)$ as a function of
deformation. All expressions are given only for the relevant case of
incompressible flows, ${\rm det}(\EE)=1$.

{\em Model 1.} This model is inspired by the work of \citet{DoiOht91} who addressed the rheology of emulsion droplets under
flow (albeit with no attempt to address the foam limit, where droplets
are in close proximity). For simplicity, Doi and Ohta assumed affine
deformation of an isotropic assembly of fluid interfaces. Their result
for $\QQ$ may be written
\begin{equation}
Q_{\mu\nu}(\EE) = -\frac{15}{4} \frac{1}{4\pi} \int \frac {
u_\mu u_\nu -
\frac{1}{3}\delta_{\mu\nu}
}{
|\EE\T\bd\uu|^4 
}\, d^2u 
\label{doexact1}
\end{equation}
where the integral is over the surface of a unit sphere; the
corresponding form for $R$ (using Eq.\ref{barriergen}) is
\begin{equation}
\frac{R(\EE)}{\lambda} = \frac{1}{4\pi}\int
\frac{d^2u}{|\EE\T\bd\uu|^4} -1.
\label{doexact2}
\end{equation}
Model 1 has $\chi = 4/15$. For small shear strains one finds
$Q_{xx}=(8/21)\gamma^2$, $Q_{yy}=-(13/21)\gamma^2$,
$Q_{zz}=(5/21)\gamma^2$ and hence $\varphi = -7/6$, while for large
strains $\varphi=-1$ as follows from $Q_{xx} = -Q_{yy}/2 = Q_{zz} =
(5/8)\gamma$. For uniaxial {\em extension} both $\QQ$ and $R$ can be
found explicitly as shown by \citet{DoiOht91}; see
Eqs.\ref{Q_uniaxial},\ref{R_uniaxial} below.

{\em Model 2.} This is a restricted simplification of Model 1, again
following \citet{DoiOht91} who presented a tractable
analytic approximant for $\QQ$ in the case of shear deformations
only. In cartesian coordinates $(x,y,z)$ with shear along $x$ and
gradient along $y$, the resulting $\QQ$ for shear strain $\gamma$
is:
\begin{equation}
\QQ(\gamma) = (1+\gamma^2/3)^{-1/2}\left(
\begin{array}{ccc}
{\gamma^2/3}&{\gamma}&{0}\\
{\gamma}&{-2\gamma^2/3}&{0}\\
{0}&{0}&{\gamma^2/3}\\
\end{array}
\right)
\label{doapprox1}
\end{equation}
and the expression for the effect of strain on stored energy,
Eq.\ref{barriergen}, gives:
\begin{equation}
\frac{R(\gamma)}{\lambda} = (1+\gamma^2/3)^{1/2} -1.
\label{doapprox2}
\end{equation}
Model 2 has $\varphi = -1$ at all strains and $\chi =1/3$.

{\em Model 3.} This is based on the work of \citet{Larson97} who,
unlike Doi and Ohta, was explicitly addressing the case of dense
foams. He modelled the fact that, in such a foam, an affine
deformation does not preserve the $120^\circ$ contact angles between
fluid films; it should therefore be followed by a very fast local
relaxation which will not show up in rheology. Larson's model for the
step strain tensor $\QQ$ reads
\begin{equation}
\QQ(\EE) = -2 \CC^{1/2} + \frac{2}{3} ({\rm tr}\,\CC^{1/2}) \mident
\label{larson1}
\end{equation}
where $\CC$ is the Cauchy strain tensor, obeying $\CC^{-1} =
\EE\T\EE$, and $\CC^{1/2}$ is the symmetric tensor with
$\CC^{1/2}\bd\CC^{1/2} = \CC$.
The resulting stored energy expression gives (via Eq.\ref{barriergen}):
\begin{equation}
\frac{R}{\lambda} = \frac{{\rm tr}\,\CC^{1/2}}{3} - 1.
\label{larson2}
\end{equation}
Model 3 has $\varphi = -3/4$ at small strains and $\varphi \to -1$ at
large ones; it predicts $\chi = 1/6$ \cite{Larson97}. These are
considerably closer to the values found numerically \cite{ReiKra00}, at
least for dry foams, than those for Models 1 and 2. This is hardly
surprising since the Doi-Ohta analysis that underlies those models
does not allow for the constancy of contact angles between films. 
(Nonetheless, for simplicity we stick to the latter for the numerics in the next section.)

We note in passing that \citet{Larson97} suggests a number
of other general forms for the tensor $\QQ$ adapted, for example,
to an assembly of elastic rather than fluid films; these could be
chosen, within the tensorial SGR framework developed here, to address
the flow of glassy assemblies of such objects.

\subsection{Numerical Results}

We now present a selection of numerical results, focussing on Model 1
and its close approximation for shear strain, Model 2. Throughout, we
work in time units such that the microscopic time scale $\Gamma_0^{-1}=1$.

The simplest case is a instantaneous step deformation $\EE$.  We
assume that before this is applied the system is in equilibrium, which
implies $x>1$ since for smaller $x$ the SGR model exhibits
aging~\cite{FieSolCat00}. From Eq.\ref{tgrce2} one then deduces, by
arguments exactly analogous to those of \citet{Sollich98}, that the
decay of the stress with time $t$ after the step deformation is given
by
\be
\sig(t) = \QQ(\EE) G_{\rm eq}(t\exp[R(\EE)/x])
\label{step_strain}
\ee
Here the time-dependence is given by the same equilibrium stress
relaxation function as in the scalar SGR model,
\be
G_{\rm eq}(z) = \frac{\int_z^\infty dz'\,G_1(z')}{\int_0^\infty dz'\,G_1(z')}
\ee
As in the scalar case, nonlinearities in the time-dependence arise
only from the factor $\exp[R(\EE)/x]$, which produces a speed-up of
the stress relaxation for large deformations. However, additional
nonlinear effects occur in the tensorial model, through the
instantaneous stress response $\QQ(\EE)$. An interesting feature of
Eq.\ref{step_strain} is that all components of the stress tensor relax
with the {\em same} time-dependence. This implies for example that the
normal stress difference ratio $\varphi=N_2/N_1$ is independent of
time, a prediction which could be tested experimentally.

\begin{figure}
\begin{center}
\epsfig{file=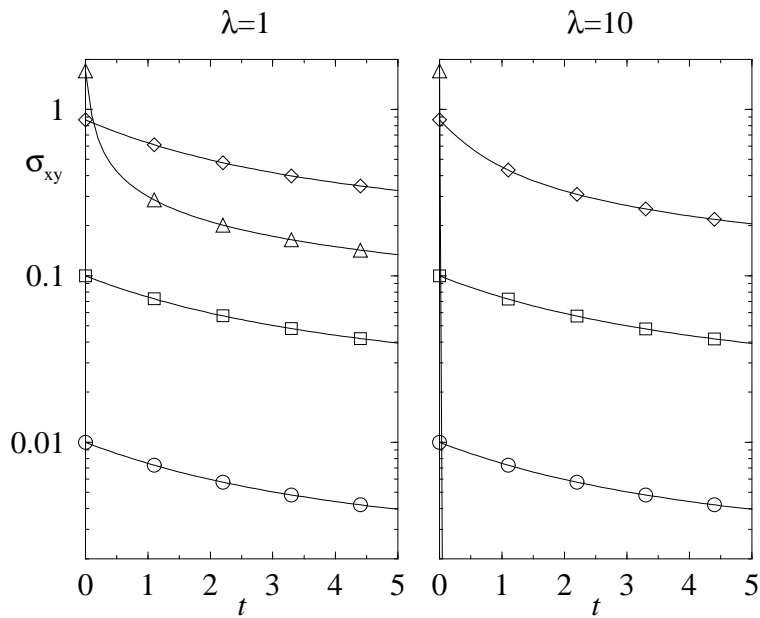,width=10cm}
\end{center}
\caption{Stress relaxation after step shear strain, calculated within
Model 2 at effective temperature $x=1.5$. Strain amplitudes are
$\gamma=0.01$ (circles), 0.1 (squares), 1 (diamonds), 10
(triangles). The two plots show the effect of variation of the
parameter $\lambda$: from Eqs.\protect\ref{barriergen}
and~\protect\ref{step_strain}, larger $\lambda$ implies greater
speed-up of the stress relaxation. For $\gamma=10$ and $\lambda=10$
(right), this speed-up is so large that the stress
relaxes to zero essentially instantaneously on the scale of the
plot. The nonlinear dependence of the initial value of the
stress on $\gamma$, which arises from the factor $\QQ(\EE)$ in
Eq.\protect\ref{step_strain}, is independent of $\lambda$ and
therefore the same in both plots.
\label{fig:step_shear}
}
\end{figure}
In figure~\ref{fig:step_shear} we show example results for step shear,
obtained from Model 2. These demonstrate the nonlinearities in both
the instantaneous stress response and the time dependence of the
subsequent stress relaxation.  Only the results for $\sigma_{xy}$ are
shown; the relaxation curves for $N_1=-N_2$ only differ through a
constant factor which reflects their different initial values.

\begin{figure}
\begin{center}
\epsfig{file=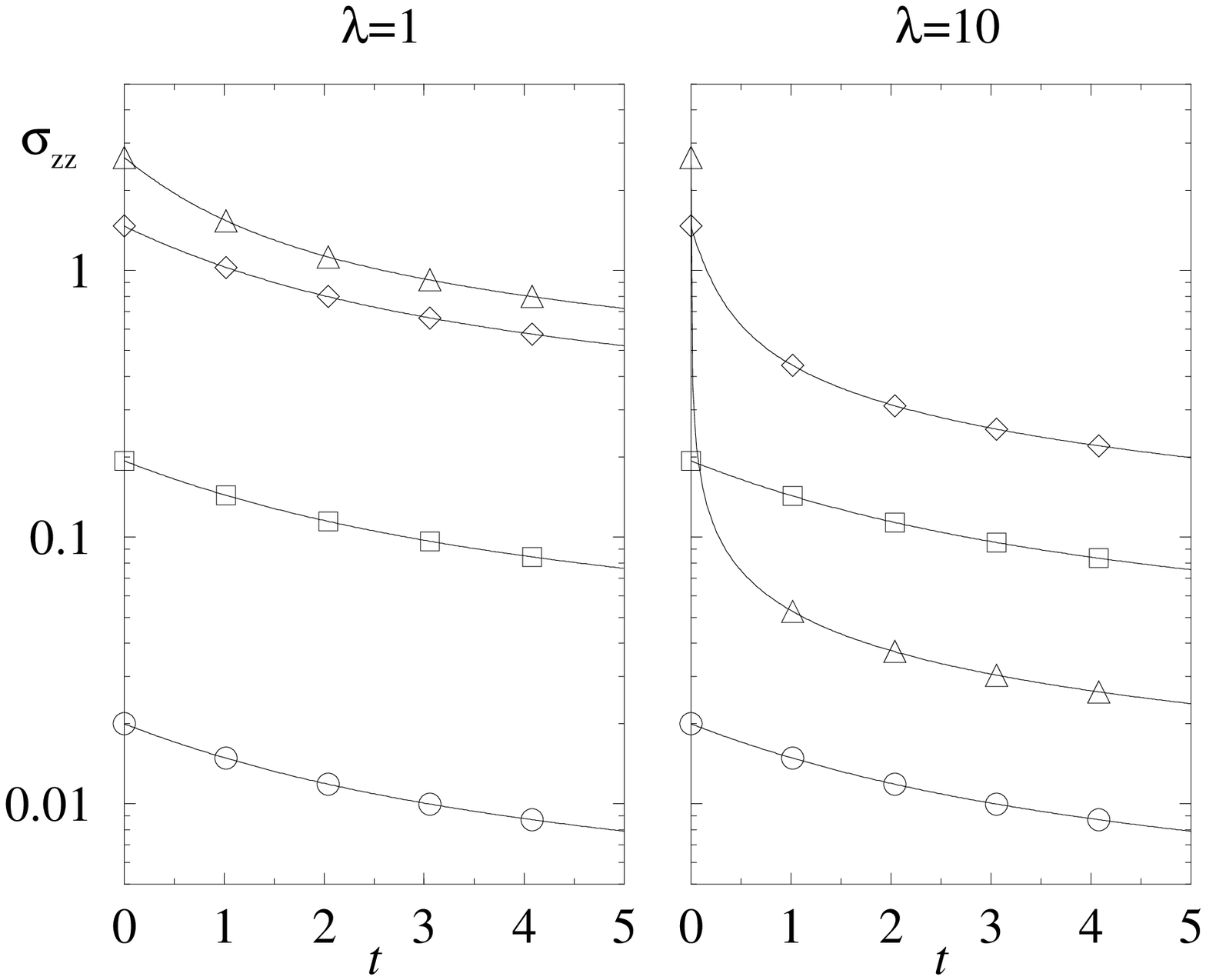,width=10cm}
\end{center}
\caption{Analogue of figure~\protect\ref{fig:step_shear} for step
uniaxial extension. Shown is the relaxation of $\sigma_{zz}$,
calculated within Model 2 at effective temperature $x=1.5$. Strain
amplitudes are $\epsilon=0.01$ (circles), 0.1 (squares), 1 (diamonds),
2 (triangles). For the smaller $\epsilon$, the initial stress response
is linear, $\sigma_{zz}=2\epsilon$, but at $\epsilon=1$ and 2 the expected
nonlinear deviations from this become apparent.
\label{fig:step_uniaxial}
}
\end{figure}
To emphasize that the tensorial SGR model can deal with deformations
other than simple shear, we show in figure~\ref{fig:step_uniaxial}
analogous results for uniaxial extension, with deformation tensor
\be
\EE = \left(\begin{array}{ccc}
e^{-\epsilon/2} & 0 & 0 \\
0 & e^{-\epsilon/2} & 0\\
0 & 0 & e^{\epsilon} \\
\end{array}
\right)
\ee
In this case we can use directly the unapproximated Model 1 since the
integrals over $\uu$ in Eqs.\ref{doexact1},\ref{doexact2} can be
performed analytically \cite{DoiOht91}. The stress tensor $\QQ(\EE)$ is
diagonal, with $Q_{xx}=Q_{yy}=-2Q_{zz}$ and\footnote{
In~\citet{DoiOht91}, the prefactor $(e^{3\epsilon}-1)^{-1}$ appears to
have been omitted from the second term in the square brackets of
Eq.\ref{Q_uniaxial}, presumably due to a
typographical error. For $\epsilon<0$ and hence $a<0$, the function
$\tau(\cdot)$ is defined by its natural analytic continuation,
$\tau(a)={\rm artanh}(\sqrt{-a})/\sqrt{-a}$.
}
\be
Q_{zz} = \frac{5}{8(e^{3\epsilon}-1)}
\left[
e^{2\epsilon}+2e^{-\epsilon} +
(e^{5\epsilon}-4e^{2\epsilon}) \,\tau\!(e^{3\epsilon}-1)
\right], \quad
\tau(a)\equiv \frac{\arctan \sqrt{a}}{\sqrt{a}}
\label{Q_uniaxial}
\ee
while the effect of deformation on stored energy is given by
\begin{equation}
\frac{R}{\lambda} = \frac{1}{2}\left[
e^{-\epsilon} + e^{2\epsilon}\,\tau\!(e^{3\epsilon}-1)
\right] - 1
\label{R_uniaxial}
\end{equation}

Next we consider steady shear flow, with shear rate $\gamdot$. From
Eq.\ref{tgrce2} one finds that the stress tensor in the steady state
is given by
\be
\sig = \frac{\int_0^\infty d\gamma\, \QQ(\gamma) \,
G_1\!\left(\gamdot^{-1}\int_0^\gamma d\gamma'\, e^{R(\gamma')/x}\right)}
{\int_0^\infty d\gamma\,
G_1\!\left(\gamdot^{-1}\int_0^\gamma d\gamma'\, e^{R(\gamma')/x}\right)}
\label{flow_curve}
\ee
where $\QQ(\gamma)$ and $R(\gamma')$ are $\QQ$ and $R$ evaluated for a shear
deformation with strain $\gamma$ and $\gamma'$, respectively. As in the
scalar SGR model, one deduces that for sufficiently large $x$ the
integrals over $\gamma$ are dominated by values $\gamma\propto
\gamdot/\Gamma_0 = \gamdot$ (recall that our underlying relaxation rate has been set to
$\Gamma_0=1$). This then gives Newtonian behaviour at small shear
rates, with $\sigma_{xy} \propto\gamdot$ and $N_1$, $N_2\propto
\gamdot^2$. As $x$ approaches the glass transition at $x=1$, however,
values of $\gamma\gg\gamdot/\Gamma_0$ remain important for small
$\gamdot$. These cause non-Newtonian singularities in the
low-$\gamdot$ behaviour,
\be
\sigma_{xy} \propto \left\{\begin{array}{ll}
\gamdot^{x-1} & \mbox{for\ }1<x<2 \\
\mbox{const.} & \mbox{for\ }x<1 \end{array} \right.
\qquad
N_1, N_2 \propto \left\{\begin{array}{ll}
\gamdot^{x-1} & \mbox{for\ }1<x<3 \\
\mbox{const.} & \mbox{for\ }x<1 \end{array} \right.
\label{steady}
\ee
These power laws, and the yield stress behaviour for $x<1$, are
largely independent of the particular forms of $\QQ$ and $R$. They
only rely on $\sigma_{xy}\propto \gamma$ and $N_1$,
$N_2\propto\gamma^2$ for small $\gamma$, and on the growth of
$R(\gamma)$ with $\gamma$ eventually limiting the largest element
strains that occur in the steady state. An interesting observation is
that the non-Newtonian effects in the normal stress differences
manifest themselves in a larger region above the glass transition (up
to $x=3$) than for the shear stress (up to
$x=2$). Another notable feature is that, when both are anomalous ($x\le 2$), normal and shear stresses obey the {\em same} power law, in contrast to the familiar analytic case ($\sigma \sim \gamdot, N \sim \gamdot^2$).
\begin{figure}
\begin{center}
\epsfig{file=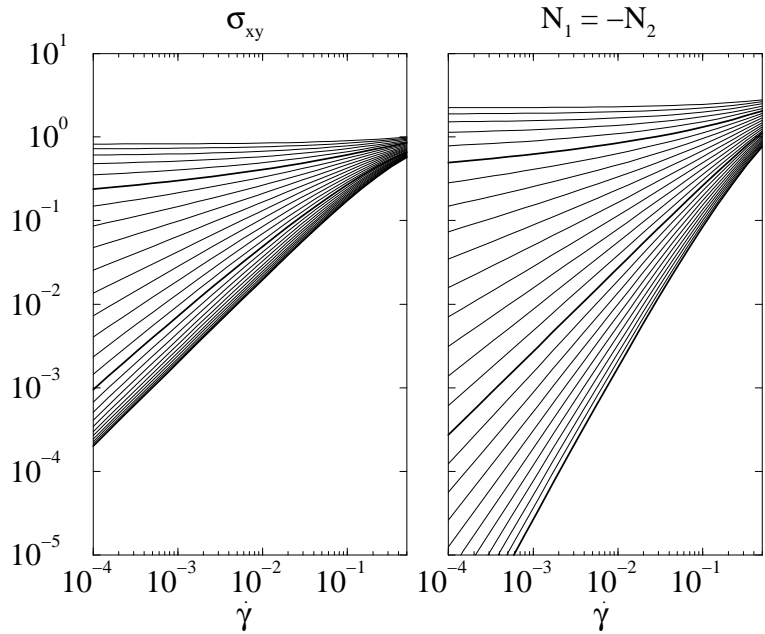,width=10cm}
\end{center}
\caption{Results for steady shear, showing shear stress $\sigma_{xy}$
(left) and normal stress differences (right) as a function of shear
rate $\gamdot$, as calculated within Model 2 with $\lambda=1$. The
curves from top to bottom are for a series of increasing noise
temperatures $x=0.5, 0.6, \ldots 3$, with $x=1$ and $x=2$ highlighted
in bold. Note that, for small $\gamdot$, the shear stress shows
Newtonian behaviour 
$\sigma_{xy}\propto \gamdot$ for $x>2$, while for the normal stress
differences non-Newtonian behaviour persists up to $x=3$. For $x<1$
the curves tend to nonzero limits for $\gamdot\to 0$, demonstrating
that the system exhibits a (dynamic) yield stress.
\label{fig:steady_shear}
}
\end{figure}
Figure~\ref{fig:steady_shear} shows some example results, calculated
from Model 2 with $\lambda=1$. The crossover from yield stress
behaviour for $x<1$ to non-Newtonian power laws for $x>1$ to Newtonian
flow for $x>2$ or $x>3$ can clearly be seen. For larger $\lambda$,
strain-induced yielding is more pronounced. This has two effects: it
limits the element strains $\gamma$ in steady state, so that at given
$\gamdot$ both $\sigma_{xy}$ and $N_1$ decrease with increasing
$\lambda$. It also means that nonlinearities caused by strain-induced
yielding are stronger, and deviations from the small-$\gamdot$
behaviour of Eq.\ref{steady} therefore appear for smaller
$\gamdot$. Numerical results for $\lambda=10$ (not shown) confirm
these expectations.

Finally, we consider a somewhat more complicated scenario, namely
stress relaxation after cessation of a steady flow. From
Eq.\ref{tgrce2} one finds for this case that the stress at a time $t$
after cessation of the flow is
\be
\sig(t) = \frac{\int_0^\infty d\gamma\, \QQ(\gamma) 
G_1\!\left(t+\frac{1}{\gamdot}\int_0^\gamma d\gamma'\, e^{R(\gamma')/x}\right)}
{\int_0^\infty d\gamma\, 
G_1\!\left(\frac{1}{\gamdot}\int_0^\gamma d\gamma'\, e^{R(\gamma')/x}\right)}
\label{cessation}
\ee
For $t=0$, this just gives the steady state shear of
Eq.\ref{flow_curve}. Eqs.\ref{flow_curve},\ref{cessation} generalize
straightforwardly to other steady flows (and their cessation); e.g.\
for an extensional flow one merely replaces $\gamma$ and $\gamdot$ by
$\epsilon$ and $\dot\epsilon$ everywhere.

We now give a brief scaling analysis of the behaviour predicted by
Eq.\ref{cessation}, focussing on the regime where all timescales
($\gamdot^{-1}$ and $t$) are large compared to the microscopic
timescale $\Gamma_0^{-1}$, which equals unity in our chosen units. The
reasoning is similar to that used for other rheological predictions of
the (scalar) SGR model \cite{Sollich98,FieSolCat00}. One uses the
following facts. The function $G_1$ has the initial value $G_1(0)=1$;
for $z\gg1$ it decays as $G_1(z)\sim t^{-x}$. The integral
$\int_0^\gamma d\gamma'\, e^{R(\gamma')/x}$ is $\approx \gamma$ for
small $\gamma$ but eventually grows very quickly with $\gamma$, in
fact typically exponentially since $R(\gamma')$ grows at least as a
power of $\gamma'$. In the denominator of Eq.\ref{cessation}, one can
thus limit the $\gamma$-integration to the $O(1)$ range before this
exponential cutoff sets in. Separating the regimes where $\gamma\leq
O(\gamdot)$ (and thus the argument of $G_1$ is $O(1)$) and $\gamma\geq
O(\gamdot)$ (where the asymptotic power law decay of $G_1$ is a good
approximation), one finds that the denominator scales as $\gamdot$ for
$x>1$ and as $\gamdot^{x}$ for $x<1$. In the numerator of
Eq.\ref{cessation} one can use a similar analysis, though the
small-$\gamma$ regime does not need to be treated separately since for
$t\gg 1$ the argument of $G_1$ is always large. For $\gamdot t\ll 1$,
the integral can again be cut off at $\gamma=O(1)$; depending on the
value of $x$, it can be dominated by the regime of small
$\gamma=O(\gamdot)$ and one then needs to bear in mind that
generically $Q_{xy}(\gamma)\sim\gamma$ while normal stress differences
such as $Q_{xx}(\gamma)-Q_{yy}(\gamma)$ scale as $\gamma^2$ for small
$\gamma$. For $\gamdot t\gg 1$, on the other hand, the cutoff in
$\gamma$ is located where $\gamdot t\approx \int_0^\gamma d\gamma'\,
e^{R(\gamma')/x}$ and so typically grows logarithmically with $\gamdot
t$; in the latter case strains $\gamma$ of $O(1)$ -- up to the cutoff --
always dominate the integral. Putting these elements together, one
finds for the relaxation of the shear stress
\be
\frac{\sigma_{xy}(t)}{\sigma_{xy}(0)} \simeq
\begin{array}{ll}
\left\{\begin{array}{p{1.8cm}l}
$1$                & \mbox{for } t\ll \gamdot^{-1} \\
$(\gamdot t)^{-x}$ & \mbox{for } t\gg \gamdot^{-1}
\end{array} \right\} & \mbox{\ for } x<2
\\
\left\{\begin{array}{p{1.8cm}l}
$t^{2-x}$                 & \mbox{for } t\ll \gamdot^{-1} \\
$t^{2-x}(\gamdot t)^{-2}$ & \mbox{for } t\gg \gamdot^{-1}
\end{array} \right\} & \mbox{\ for } x>2
\end{array}
\label{shear_cess}
\ee
and for a typical normal stress difference such as $N_1$
\be
\frac{N_1(t)}{N_1(0)} \simeq
\begin{array}{ll}
\left\{\begin{array}{p{1.8cm}l}
$1$                & \mbox{for } t\ll \gamdot^{-1} \\
$(\gamdot t)^{-x}$ & \mbox{for } t\gg \gamdot^{-1}
\end{array} \right\} & \mbox{\ for } x<3
\\
\left\{\begin{array}{p{1.8cm}l}
$t^{3-x}$                 & \mbox{for } t\ll \gamdot^{-1} \\
$t^{3-x}(\gamdot t)^{-3}$ & \mbox{for } t\gg \gamdot^{-1}
\end{array} \right\} & \mbox{\ for } x>3
\end{array}
\label{norm_cess}
\ee
In Eqs.\ref{shear_cess},\ref{norm_cess} the power laws for
$t\gg\gamdot^{-1}$ are all subject to logarithmic corrections in
$\gamdot t$, which arise from the variation of the $\gamma$-cutoff
discussed above. The scalings given apply to the generic tensorial SGR
model of Eq.\ref{tgrce2}, including in particular our Models 1--3. The
glassy nature of the model is manifested in the relaxations scaling
with $\gamdot t$ for low enough $x$, rather than with $t$ as one would
expect if the microscopic timescale dominates the dynamics. As in the
results for steady shear, Eq.\ref{steady}, glassy effects remain
important up to higher $x$ for normal stress differences ($x=3$) than
for the shear stress ($x=2$).

\begin{figure}
\begin{center}
\epsfig{file=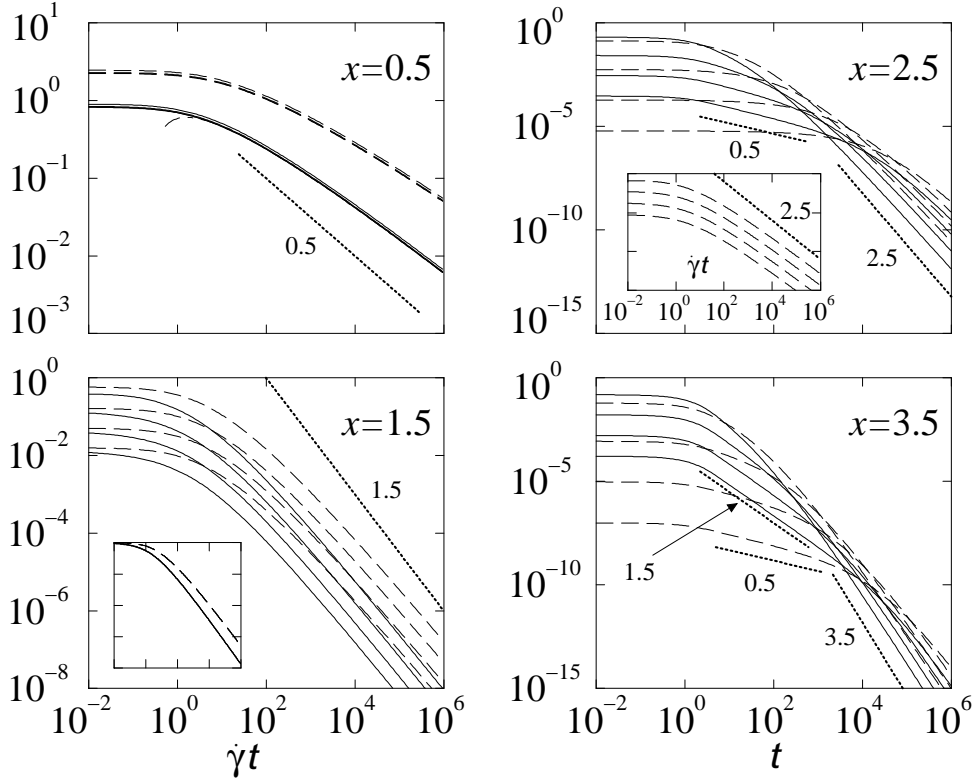,width=13cm}
\end{center}
\caption{Results for cessation of steady shear flow, calculated within
Model 2 with $\lambda=1$, for four different noise temperatures $x$. Shear
stress $\sigma_{xy}$ and first normal stress difference $N_1=-N_2$ are
plotted as solid and dashed lines, respectively; in each graph a set
of four curves are shown for each quantity, for shear rates
$\gamdot=10^{-4}, 10^{-3}, 10^{-2}, 10^{-1}$ from bottom to
top. Theoretically predicted power laws from
Eqs.\protect\ref{shear_cess},\protect\ref{norm_cess} are shown by bold
dotted lines and annotated with their (negative) slopes. Top left: For
$x=0.5$; both $\sigma_{xy}$ and $N_1$ scale with $\gamdot
t$. Logarithmic corrections to the predicted decay $\sim(\gamdot
t)^{-x}$ are non-negligible in the range of $\gamdot t$ shown. The
underlying power law is nevertheless correct, as a fit of
$\sigma_{xy}(t)$ for $\gamdot=10^{-4}$ to the form $[a\ln (\gamdot
t)+b](\gamdot t)^{-1/2}$ shows (dashed-dotted line, just
distinguishable around $\gamdot t=1$). Bottom left: $x=1.5$; the larger
decay exponent $(\gamdot t)^{-x}$ now makes logarithmic corrections
less important. In the inset the curves are rescaled vertically by
their value at $t=0$ to show that their time dependence is otherwise
identical. Top right: $x=2.5$. The shear stress $\sigma_{xy}$ follows
the predicted power laws from Eq.\protect\ref{shear_cess} quite
well. In constrast to $\sigma_{xy}$, the normal stress difference only
decays significantly for $t\gg\gamdot^{-1}$, scaling throughout with
$\gamdot t$. This is shown in the inset where (on the same vertical
scale) $N_1$ is plotted against $\gamdot t$. Bottom right: for
$x=3.5$, both $\sigma_{xy}$ and $N_1$ decay for $t\ll\gamdot^{-1}$.
\label{fig:cessation}
}
\end{figure}
In figure~\ref{fig:cessation} we show some numerical results,
calculated from Eq.\ref{cessation} within Model 1 with $\lambda=1$.
The power laws predicted by Eqs.\ref{shear_cess},\ref{norm_cess} are
quite well obeyed, subject to the expected logarithmic corrections for
$t\gg\gamdot^{-1}$. As a general trend, we note that the normal stress
difference $N_1$ decays more slowly than the shear stress
$\sigma_{xy}$. This is certainly true for $x>3$, where $N_1$ decays
with a slower power law than $\sigma_{xy}$ from
Eqs.\ref{shear_cess},\ref{norm_cess}, and for $2<x<3$ where $N_1$
remains essentially constant for $t\ll\gamdot^{-1}$. From
figure~\ref{fig:cessation} we also observe the same trend for $x<2$,
however: even though Eqs.\ref{shear_cess},\ref{norm_cess} predict that
$\sigma_{xy}$ and $N_1$ should have the same asymptotic decay shapes
in this regime, $N_1$ is seen to decay more slowly than $\sigma_{xy}$,
presumably due to stronger logarithmic corrections.

\section{Conclusions}
In this work we have presented a general strategy for alleviating one
of the more serious restrictions of the SGR model for soft glasses,
namely its effectively scalar treatment of stress and strain, which
results in an inability to consider normal stresses or flows other
than simple shear. As might be expected, the tensorialization allows
significant scope for tailoring the model to different classes of soft
material, without altering the rather simple basic assumptions that
underlie it. In particular, we have presented results for several
tensorial SGR variants that offer a possible description of flowing
foams and/or dense emulsions. Because of the amorphous packings and
slow dynamics observed in these materials, they are prime candidates
for the physical picture that underlies the SGR model. It is useful,
we believe, to have a constitutive equation that can capture to
reasonable accuracy the step-strain response of real foams (Model 3 is
probably best in this respect) while also predicting nontrivial yield
behavior, flow curves, and aging phenomena. These three features all
arise in the glass phase of the model ($x\le 1$) which is therefore
the one most likely to be of interest in foam flows, except perhaps
very close to the onset of rigidity which occurs at volume fractions
of the dispersed phase of around 58\% \cite{MasBibWei96}. 

As emphasized elsewhere \cite{SolLeqHebCat97,Sollich98,FieSolCat00},
the absence of tensorial structure is by no means the only shortcoming
of the scalar SGR model; its various assumptions are all questionable
at several levels and, even after tensorialization, the approach
should not be viewed as complete in any sense. Experimental
falsifications of its predictions are welcome, since these will help
direct theoretical work towards more complete models. (Verifications are, of course, also
welcome.) Our development in this paper of tensorial versions of SGR
should certainly allow more stringent comparisons to be drawn between
theory and experiments.

Indeed, in rheological terms one could argue that a proper tensorial
treatment of stress and strain is the absolute minimum required for
any kind of serious predictive modelling to begin. In this sense, the
models presented in this paper could represent a `coming of age' for
the SGR approach to the rheology and rheological aging of soft
materials. Despite the shortcomings of SGR, we are not aware of any
competing approaches that directly confront the nonergodic features of
these systems within a rheological context.

\subsection*{Acknowledgements} We acknowledge useful discussions with
M Doi, A Kraynick, R H\"{o}hler, and C Holmes. We thank the Newton Institute,
where part of this work was done, for hospitality.


\begin{thebibliography}{20}
\expandafter\ifx\csname natexlab\endcsname\relax\def\natexlab#1{#1}\fi

\bibitem[Bouchaud(1992)]{Bouchaud92}
J~P Bouchaud.
\newblock Weak ergodicity breaking and aging in disordered-systems.
\newblock {\em J.\ Phys.\ (France)\ I} {\bf 2}, 1705--1713 (1992).

\bibitem[Bouchaud et~al.(1995)Bouchaud, Comtet, and Monthus]{BouComMon95}
J~P Bouchaud, A~Comtet, and C~Monthus.
\newblock On a dynamical model of glasses.
\newblock {\em J.\ Phys.\ (France)\ I} {\bf 5}, 1521--1526 (1995).

\bibitem[Cates(2003)]{Cates_LesHouches03}
M~E Cates.
\newblock Structural relaxation and rheology of soft condensed matter.
\newblock In J~L Barrat and J~Kurchan, editors, {\em Slow Relaxations and
  Nonequilibrium Dynamics in Condensed Matter (77th Les Houches Summer
  School)}. Springer, 2003.

\bibitem[Clo{\^{\i}}tre et~al.(2000)Clo{\^{\i}}tre, Borrega, and
  Leibler]{CloBorLei00}
M~Clo{\^{\i}}tre, R~Borrega, and L~Leibler.
\newblock Rheological aging and rejuvenation in microgel pastes.
\newblock {\em Phys.\ Rev.\ Lett.} {\bf 85}, 4819--4822 (2000).

\bibitem[Clo{\^{\i}}tre et~al.(2003)Clo{\^{\i}}tre, Borrega, Monti, and
  Leibler]{CloBorMonLei03}
M~Clo{\^{\i}}tre, R~Borrega, F~Monti, and L~Leibler.
\newblock Glassy dynamics and flow properties of soft colloidal pastes.
\newblock {\em Phys.\ Rev.\ Lett.} {\bf 90} 068303 (2003).

\bibitem[Cohen-Addad and H{\"{o}}hler(2001)]{CohHoh01}
S~Cohen-Addad and R~H{\"{o}}hler.
\newblock Bubble dynamics relaxation in aqueous foam probed by multispeckle
  diffusing-wave spectroscopy.
\newblock {\em Phys.\ Rev.\ Lett.} {\bf 86}, 4700--4703 (2001).

\bibitem[Doi and Ohta(1991)]{DoiOht91}
M~Doi and T~Ohta.
\newblock Dynamics and rheology of complex interfaces. 1.
\newblock {\em J.\ Chem.\ Phys.} {\bf 95}, 1242--1248 (1991).

\bibitem[Fielding et~al.(2000)Fielding, Sollich, and Cates]{FieSolCat00}
S~M Fielding, P~Sollich, and M~E Cates.
\newblock Aging and rheology in soft materials.
\newblock {\em J.\ Rheol.} {\bf 44}, 323--369 (2000).

\bibitem[Head et~al.(2001)Head, Ajdari, and Cates]{HeaAjdCat01}
D~A Head, A~Ajdari, and M~E Cates.
\newblock Jamming, hysteresis, and oscillation in scalar models for shear
  thickening.
\newblock {\em Phys.\ Rev.\ E} {\bf 64}, 061509 (2001).

\bibitem[Head et~al.(2002)Head, Ajdari, and Cates]{HeaAjdCat02}
D~A Head, A~Ajdari, and M~E Cates.
\newblock Rheological instability in a simple shear-thickening model.
\newblock {\em Europhys.\ Lett.}, {\bf 57}, 120--126 (2002).

\bibitem[H{\"{o}}hler et~al.(1999)H{\"{o}}hler, Cohen-Addad, and
  Asnacios]{HohCohAsn99}
R~H{\"{o}}hler, S~Cohen-Addad, and A~Asnacios.
\newblock Rheological memory effect in aqueous foam.
\newblock {\em Europhys.\ Lett.} {\bf 48}, 93--98 (1999).

\bibitem[Holdsworth(1993)]{Holdsworth93}
S~D Holdsworth.
\newblock Rheological models used for the prediction of the flow properties of
  food products.
\newblock {\em Trans.\ Inst.\ Chem.\ Eng.} {\bf 71}, 139--179 (1993).

\bibitem[Larson(1997)]{Larson97}
R~G Larson.
\newblock The elastic stress in ``film fluids''.
\newblock {\em J.\ Rheol.} {\bf 41}, 365--372 (1997).

\bibitem[Mason et~al.(1996)Mason, Bibette, and Weitz]{MasBibWei96}
T~G Mason, J~Bibette, and D~A Weitz.
\newblock Yielding and flow of monodisperse emulsions.
\newblock {\em J.\ Coll.\ Interf.\ Sci.} {\bf 179}, 439--448 (1996).

\bibitem[Michel et~al.(2001)Michel, Appell, Molino, Kieffer, and
  Porte]{MicAppMolKiePor01}
E~Michel, J~Appell, F~Molino, J~Kieffer, and G~Porte.
\newblock Unstable flow and nonmonotonic flow curves of transient networks.
\newblock {\em J.\ Rheol.} {\bf 45}, 1465--1477 (2001).

\bibitem[Reinelt and Kraynik(2000)]{ReiKra00}
D~A Reinelt and A~M Kraynik.
\newblock Simple shearing flow of dry soap foams with tetrahedrally
  close-packed structure.
\newblock {\em J.\ Rheol.} {\bf 44}, 453--471 (2000).

\bibitem[Sollich(1998)]{Sollich98}
P~Sollich.
\newblock Rheological constitutive equation for a model of soft glassy
  materials.
\newblock {\em Phys.\ Rev.\ E} {\bf 58}, 738--759 (1998).

\bibitem[Sollich et~al.(1997)Sollich, Lequeux, H{\'{e}}braud, and
  Cates]{SolLeqHebCat97}
P~Sollich, F~Lequeux, P~H{\'{e}}braud, and M~E Cates.
\newblock Rheology of soft glassy materials.
\newblock {\em Phys.\ Rev.\ Lett.} {\bf 78}, 2020--2023 (1997).

\bibitem[Viasnoff et~al.(2003)Viasnoff, Jurine, and Lequeux]{ViaJurLeq02}
V~Viasnoff, S~Jurine, and F~Lequeux.
\newblock How colloidal suspensions that age are rejuvenated by strain
  application.
\newblock {\em Faraday Discussions} {\bf 123}, 253--266 (2003).

\bibitem[Viasnoff and Lequeux(2002)]{ViaLeq02}
V~Viasnoff and F~Lequeux.
\newblock Rejuvenation and overaging in a colloidal glass under shear.
\newblock {\em Phys.\ Rev.\ Lett.} {\bf 89}, 065701 (2002).

\end{thebibliography}

\end{document}